\documentclass[journal,10pt]{IEEEtran}

\usepackage{xcolor}
\usepackage{relsize}

\usepackage{ifpdf}
\usepackage{graphicx}
\usepackage{cite}
\usepackage{nomencl}
\usepackage{enumitem}
\usepackage[normalem]{ulem}

\usepackage{amsmath,amssymb,bm,bbm}
\usepackage{algorithmic}
\usepackage{adjustbox}
\usepackage{array}
\usepackage[tablename=Table]{caption}
\usepackage{stfloats}
\usepackage{url}
\usepackage{amsthm}
\usepackage{multirow}
\usepackage{optidef}
\usepackage{dirtytalk}

\usepackage{stackengine,scalerel}

\newcommand\overstar[1]{\ThisStyle{\ensurestackMath{%
			\setbox0=\hbox{$\SavedStyle#1$}%
			\stackengine{0pt}{\copy0}{\kern.2\ht0\smash{\SavedStyle*}}{O}{c}{F}{T}{S}}}}

\begin{document}
 
\title{A Sufficient Condition to Guarantee Non-Simultaneous Charging and Discharging of Household Battery Energy Storage}

\author{\IEEEauthorblockN{Amit Joshi, Hamed Kebriaei,~\textit{Senior Member, IEEE}, \\ Valerio Mariani,~\textit{Member, IEEE}, Luigi Glielmo,~\textit{Senior Member, IEEE}
        \thanks{A. Joshi, V. Mariani and L. Glielmo are with the Department of Engineering, Università del Sannio, 82100 Benevento, Italy, (e-mail: amit.joshi@unisannio.it; valerio.mariani@unisannio.it; glielmo@unisannio.it)}
		\thanks{H. Kebriaei is with the School of Electrical and Computer Engineering, College of Engineering, University of Tehran, 1417466191 Tehran, Iran, (e-mail: kebriaei@ut.ac.ir).}
		\thanks{*It has received funding from the Italian Ministry of University and Research PRIN 2017 project VECTORS}
		}}

\markboth{Preprint submitted to IEEE PES Letters}%
{Sai \MakeLowercase{\textit{et al.}}: Bare Demo of IEEEtran.cls for IEEE Journals}

\maketitle

\begin{abstract}
In this letter, we model the day-ahead price-based demand response of a residential household with battery energy storage and other controllable loads, as a convex optimization problem. Further using duality theory and Karush-Kuhn-Tucker optimality conditions, we derive a sufficient criterion which guarantees non-simultaneous charging and discharging of the battery energy storage, without explicitly modelling it as a constraint.
\end{abstract}

\vspace{-0.5cm}
\section{Introduction}
\IEEEPARstart{I}{ntergration} of battery energy storage (BES) along with renewable energy resources (RERs), adds to the viability of residential demand response (DR), due to increased system flexibility \cite{haider2016review}. The state-of-art DR programs \cite{albadi2007demand}, model the BES using variants of the energy reservoir model, which differ in terms of charging/discharging efficiencies and self-discharge power \cite{rosewater2018battery}. However, irrespective of the model used, one has to ensure that the battery cannot be charged and discharged simultaneously, also known as \say{non-simultaneousness} or \say{complementarity} constraint.

The complementarity constraint makes the problem non-convex and standard relaxation techniques have been used to ensure non-simultaneousness, using the convex equivalent. In this regard, \cite{zarrilli2017energy} proposes a penalty based relaxation to discourage non-simultaneousness. The authors in \cite{garifi2020convex} provide sufficient conditions under which the penalty based approach ensures non-simultaneousness. A similar study was done for a residential household with linear penalty term \cite{garifi2018control}. 

In comparison to \cite{zarrilli2017energy, garifi2018control, garifi2020convex}, the novelties of our work are: 1) the purchasing and selling tariff are different, resulting in a non-smooth function; 2) we model additional controllable loads such as thermostats, washing machine, etc.; 3) we derive sufficient conditions based on charging/discharging efficiencies, to ensure non-simultaneousness, thereby eliminating the need of penalty based relaxation.

The rest of the paper is organized as follows: in Section II, we model the controllable loads; the relaxed optimization problem and the sufficient condition to ensure non-simultaneousness is stated in Section III; the simulation studies are described in Section IV.
\vspace{-.3cm}
\section{Modelling}
We consider a residential household participating in a horizon-ahead, price-based, demand response program \cite{albadi2007demand}. The household is equipped with a household energy management system (HEMS), which responds to the horizon-ahead electricity price by adjusting the controllable loads, and consequently coordinating the energy exchange of the grid and the renewable energy resources (RERs).

Let $\mathcal{K} = \{ 0,\dots,k,\dots, K-1 \}$ denote the time horizon, with $k$ as the discrete time index and $\Delta t$ as the discretization step. Let $r_{k}$ and $g_{k}$ represent the renewable energy generation and the energy exchange with the grid, respectively; the purchasing/selling price from/to the grid is denoted by $p_{k}$ and $s_{k}$, respectively.

\vspace{-.5cm}
\subsection{Critical and Controllable Loads}
The appliances of the house contribute to the aggregate electrical load, and can be classified as critical or controllable loads. The critical loads, denoted as $d_{k}$, indicate the demand that must be met at all times. On the other hand, the controllable loads indicate the demand that can be adjusted, in response to the electricity tariff.
In the subsequent, we model two variants of controllable loads, namely \say{dynamical} and \say{non-dynamical}.
\subsubsection{Dynamical Controllable Loads}
The controllable loads whose behaviour is governed by dynamical equations, such as BES and thermostatically controllable loads (TCL). A 
BES can represent a conventional Li-ion battery or an electric vehicle (EV). Likewise, a TCL can represent an air conditioner or a heater.\\
\textbf{BES:} We model the BES using the energy reservoir model proposed in \cite{rosewater2018battery} and \cite{parisio2014model}.
Let $u_{k}^{\rm{ch}}$ and $u_{k}^{\rm{dch}}$ be the control variables of the battery, which respectively indicate the charging power and the discharging power, and must satisfy
\begin{subequations}
 \begin{align}
  0 \leq u_{k}^{\rm{ch}} \leq \overline{u}^{\rm{ch}},
  \label{eqn:Battery_charging_constraint}\\
  0 \leq u_{k}^{\rm{dch}} \leq \overline{u}^{\rm{dch}},
  \label{eqn:Battery_discharging_constraint}\\
  u_{k}^{\rm{ch}}u_{k}^{\rm{dch}} = 0,
  \label{eqn:battery_non_simultaneous_constraint}
 \end{align}
\label{eqn:battery_constraints}%
\end{subequations}
$\forall k \in \mathcal{K}$. In (\ref{eqn:battery_constraints}), $\overline{u}^{\rm{ch}}$ and $\overline{u}^{\rm{dch}}$ indicate the maximum charging and discharging power, respectively; (\ref{eqn:battery_non_simultaneous_constraint}) ensures that the battery cannot be charged and discharged simultaneously.\\
Let $x_{k}$ be the normalized state-of-charge (SoC) of the battery, whose dynamics are governed as 
\begin{equation}
    x_{k+1} = x_{k} + \Delta t \big( \eta^{\rm{ch}} u_{k}^{\rm{ch}} - \frac{1}{\eta^{\rm{dch}}} u_{k}^{\rm{dch}} - u^{\rm{sd}}\big)/{E},
    \label{eqn:SoC_dynamics}
\end{equation}
where $x_{0}$ is the given initial SoC, $u^{\rm{sd}}$ is the self-discharge power, $E$ is the energy capacity of the battery, $\eta^{\rm{ch}}$ and $\eta^{\rm{dch}}$ are the charging and discharging efficiencies of the battery, such that $0 < \big( \eta^{\rm{ch}},\, \eta^{\rm{dch}} \big) \leq 1$.
The normalized SoC must satisfy
\begin{equation}
    \underline{x} \leq x_{k+1} \leq \overline{x}, \hspace{.5cm} \forall k \in \mathcal{K},
    \label{eqn:SoC_constraints}
\end{equation}
where $0 \leq \underline{x} < \overline{x} \leq 1$ and $x_{0} \in [\underline{x},\overline{x}]$.\\
\textbf{TCL:} We model the TCL using the discrete-time equivalent of the continuous-time model proposed in \cite{hao2014aggregate}. Let $\theta_{k}$ be the indoor temperature, $\theta_{k}^{\rm{ex}}$ be the external temperature profile\footnote{We assume that all the exogenous inputs, i.e., occupancy, solar disturbance, etc., have been perfectly captured by~$\boldsymbol{\theta}^{\rm{ex}}$.} and $\theta^{\rm{s}}$ be the given set-point temperature; let $\Delta^{\rm{d}} > 0$ be the dead-band. The global thermal capacitance, thermal resistance and coefficient of performance are respectively denoted by $C$, $R$ and $\rm{CoP}$.\\
Let $u_{k}^{\rm{tcl}}$ be the control variable for the TCL which must satisfy
\begin{equation}
    0 \leq u_{k}^{\rm{tcl}} \leq \overline{u}^{\rm{tcl}}, \hspace{.5cm} \forall k \in \mathcal{K},
    \label{eqn:AC_charging_constraint}
\end{equation}
where $\overline{u}^{\rm{tcl}}$ is its rated capacity. The control variable influences the indoor temperature over time, described as 
\begin{equation}
    \theta_{k+1} = \Tilde{a} \theta_{k} + \Delta t( a\theta^{\rm{ex}} - b u_{k}^{\rm{tcl}}),
    \label{eqn:temperature_dynamics}
\end{equation}
where $\theta_{0}$ is the given initial temperature, $\Tilde{a} = 1-a\Delta t$, $a = 1/RC$ and $b = {\rm{CoP}}/{C}$.\\
Further, the indoor temperature at all times has to stay within an acceptable range, given as
\begin{equation}
    \underline{\theta} \leq \theta_{k+1} \leq \overline{\theta}, \hspace{.5cm} \forall k \in \mathcal{K},
    \label{eqn:temperature_constraints}
\end{equation}
where $\underline{\theta} = \theta^{\rm{s}} - \Delta^{\rm{d}}$, $\overline{\theta} = \theta^{\rm{s}} + \Delta^{\rm{d}}$ and $\theta_{0}\in [\underline{\theta},\overline{\theta}]$
\subsubsection{Non-dynamical controllable loads}
The energy consumption of such loads at different time instants are not coupled via dynamical equations; washing machine and dish washer fall in this category. We model non-dynamical loads using
\begin{equation}
    \underline{u}^{\rm{nd}} \leq u_{k}^{\rm{nd}} \leq \overline{u}^{\rm{nd}}, \quad \forall k \in \mathcal{K},
    \label{eqn:constraint_deferrable_1}
\end{equation}
and
\begin{equation}
    \sum_{k \in \mathcal{K}} u_{k}^{\rm{nd}} = \mathbf{U}^{\rm{nd}},
    \label{eqn:constraint_deferrable_2}
\end{equation}
where $u_{k}^{\rm{nd}}$ is the instantaneous consumption, which can be adjusted within the bounds, as defined in (\ref{eqn:constraint_deferrable_1}).  Further, (\ref{eqn:constraint_deferrable_2}) ensures that the aggregate consumption of the dynamical load over the horizon remains unaffected.
\vspace{-.5cm}
\subsection{Electricity Bill}
The energy exchange with the grid is given by 
\begin{equation}
    g_{k} = \underbrace{u_{k}^{\rm{ch}} - u_{k}^{\rm{dch}} + u_{k}^{\rm{nd}} + u_{k}^{\rm{tcl}}}_{\text{control variables}} + d_{k} - r_{k}.
    \label{eqn:load_balance}
\end{equation}
Based on the energy exchange with the grid, the cost of electricity is given by
\begin{equation}f_{k}^{e} = 
\begin{cases}
p_{k}g_{k}, & \text{if $g_{k}  \geq 0$ (purchase from grid)}, \\
s_{k}g_{k}, & \text{otherwise (supply to grid)}.
\end{cases}
\label{eqn:electricity_cost}
\end{equation}
As for the majority of real-world applications, we assume the purchasing price is larger than the selling price and both are strictly positive, i.e., $p_{k} \geq s_{k} > 0,\, \forall k \in \mathcal{K}$. Such an assumption allows to re-write (\ref{eqn:electricity_cost}) as
\begin{equation}
    f_{k}^{e} = \text{max}\big\{ p_{k}g_{k},\, s_{k}g_{k}\big\}.
    \label{eqn:electricity_cost_modified}
\end{equation}
Note that the electricity bill in (\ref{eqn:electricity_cost_modified}) is piece-wise linear, non-differentiable at $g_{k} = 0$.
\vspace{-.5cm}
\subsection{Regularization cost}
The regularization cost models the \say{negative utility} of the associated variable and is often included as an additional cost.  Typically, such costs are modelled using quadratic and/or linear terms, given as
\begin{equation}
\begin{aligned}
        f_{k}^{r} = \underbrace{\alpha^{\rm{ch}}\big( u_{k}^{\rm{ch}}\big)^{2} + \beta^{\rm{ch}}u_{k}^{\rm{ch}}}_{\text{BES charging}} + \underbrace{\alpha^{\rm{dch}}\big( u_{k}^{\rm{dch}}\big)^{2} + \beta^{\rm{dch}}u_{k}^{\rm{dch}}}_{\text{BES discharging}} + \\ \underbrace{\alpha^{\rm{nd}} \big( u_{k}^{\rm{nd}} \big)^{2} + \beta^{\rm{nd}}\big( u_{k}^{\rm{nd}} \big)}_{\text{non-dynamical load}} + \underbrace{\alpha^{\rm{tcl}} \big( u_{k}^{\rm{tcl}} \big)^{2} + \beta^{\rm{tcl}}\big( u_{k}^{\rm{tcl}} \big)}_{\text{TCL}}
\end{aligned}
    \label{eqn:regularization_cost}
\end{equation}
where $(\alpha^{\rm{ch}},\beta^{\rm{ch}},\alpha^{\rm{dch}},\beta^{\rm{dch}},\alpha^{\rm{nd}},\beta^{\rm{nd}},\alpha^{\rm{tcl}},\beta^{\rm{tcl}})\geq 0$.
Usually (\ref{eqn:regularization_cost}) is used to penalize the energy exchange with the respective component (BES, TCL and non-dynamical loads), which also serves as a proxy for life-cycle/degradation of the equipment.
\vspace{-.3cm}
\section{Relaxed Convex Formulation}
Using (\ref{eqn:load_balance}), we express the electricity bill and the regularization cost in terms of the control variables ($u_{k}^{\rm{ch}}$,\, $u_{k}^{\rm{dch}}$,\,$u_{k}^{\rm{nd}}$ and $u_{k}^{\rm{tcl}}$). The optimization problem in the \say{standard form} can be written as
\begin{mini}|l|
  { u_{\mathcal{K}} }{ \sum_{k \in \mathcal{K} }f(u_{k}) }{}{}
  \addConstraint{ h^{i}(u_{k}) \leq 0 }, \quad \forall i \in \mathcal{I}, \forall k \in \mathcal{K}
  \addConstraint { l(u_{\mathcal{K}}) = 0 },
  \label{eqn:relaxed_optimization_problem}
\end{mini}
where $u_{k} = (u_{k}^{\rm{ch}},\,u_{k}^{\rm{dch}},\,u_{k}^{\rm{nd}},\,u_{k}^{\rm{tcl}})$, $f = f_{k}^{e} + f_{k}^{r}$ and $u_{\mathcal{K}} = \{ u_{0},\dots,u_{K-1} \}$. The index of inequalities is given by the set $\mathcal{I} = \{ {\rm{ch}\uparrow},\, {\rm{ch}\downarrow},\,{\rm{dch}\uparrow},\, {\rm{dch}\downarrow},\,{\rm{nd}\uparrow},\,
{\rm{nd}\downarrow},\,{\rm{tcl}\uparrow},\,
{\rm{tcl}\downarrow},\,{x\uparrow},\,{x\downarrow},\,\\{\theta \uparrow},\, {\theta \downarrow} \}$, where the corresponding functions (with slight abuse of notation) are $h^{\rm{ch}\uparrow} := -u_{k}^{\rm{ch}}$,\, $h^{\rm{ch}\downarrow} := u_{k}^{\rm{ch}}-\overline{u}^{\rm{ch}}$,\,$h^{\rm{dch}\uparrow} := -u_{k}^{\rm{dch}}$,\, $h^{\rm{dch}\downarrow} := u_{k}^{\rm{dch}}-\overline{u}^{\rm{dch}}$,\,$h^{\rm{nd}\uparrow} := -u_{k}^{\rm{nd}}$,\, $h^{\rm{nd}\downarrow} := u_{k}^{\rm{nd}}-\overline{u}^{\rm{ch}}$,\,$h^{\rm{tcl}\uparrow} := -u_{k}^{\rm{tcl}}$,\, $h^{\rm{tcl}\downarrow} := u_{k}^{\rm{tcl}}-\overline{u}^{\rm{tcl}}$,\,$h^{x\uparrow} := \underline{x}-x_{k+1}$,\, $h^{x\downarrow} := x_{k+1}-\overline{x}$,\,$h^{\theta \uparrow} := \underline{\theta}-\theta_{k+1}$,\, $h^{\theta\downarrow} := \theta_{k+1}-\overline{\theta}$. The equality constraint is $l(u_{\mathcal{K}}) := \sum_{k \in \mathcal{K}} u_{k}^{\rm{nd}}-\mathbf{U}$.\\
\textbf{Remark 1:} It is worth noting that the constraint (\ref{eqn:battery_non_simultaneous_constraint}) is relaxed, in particular not included, in (\ref{eqn:relaxed_optimization_problem}).\\
\textbf{Theorem 1:} The optimal battery charging and discharging profile obtained from (\ref{eqn:relaxed_optimization_problem}) satisfies the non-simultaneousness constraint, i.e. (\ref{eqn:battery_non_simultaneous_constraint}) for all $k \in \mathcal{K}$, if $\eta^{\rm{ch}}\eta^{\rm{dch}} < 1$.\\
\begin{proof}
By contradiction, let the optimal solution of (\ref{eqn:relaxed_optimization_problem}) be $u_{k}^{\rm{ch}*} \text{and}\, u_{k}^{\rm{dch}*}$, such that $u_{k}^{\rm{ch}*}u_{k}^{\rm{dch}*}\neq 0$ for some $k \in \mathcal{K}$. As the primal optimization problem is convex and the constraints are affine, from Slater's constraint qualification, strong duality holds \cite{boyd2004convex}. Consequently, the optimal primal-dual solution of (\ref{eqn:relaxed_optimization_problem}) satisfies the Karush-Kuhn-Tucker (KKT) conditions \cite{gordon2012karush}, given by
\vspace{-.1cm}
\begin{enumerate}
    \item Stationarity\\
    $0 \in \partial f(u_{k}) + \sum_{i \in \mathcal{I}} \lambda_{k}^{i} \partial h^{i}(u_{k}) + \nu {\partial l(u_{\mathcal{K}})},\, \forall k \in \mathcal{K}$;\\
    \vspace{-.3cm}
    \item Complementary slackness\\
    $\lambda_{k}^{i}h^{i}(u_{k}) = 0,\, \forall k \in \mathcal{K},\, \forall i \in \mathcal{I}$;\\
    \vspace{-.3cm}
    \item Primal feasibility\\
    $l(u_{\mathcal{K}}) = 0\, \text{and}\, h^{i}(u_{k}) \leq 0, \forall k \in \mathcal{K} ,\, \forall i \in \mathcal{I}$;\\
    \vspace{-.3cm}
    \item Dual feasibility\\
    $\lambda_{k}^{i} \geq 0, \forall k \in \mathcal{K} ,\, \forall i \in \mathcal{I}$;
\end{enumerate}
where ${\partial f(\cdot)}, \partial h^{i}(\cdot)$ and $\partial l(\cdot)$ are the sub-differential\footnote{The sub-differential is the set of sub-gradients which is non-empty even if the function if not differentiable, and convex.} sets of $f, h^{i}$ and $l$ w.r.t. $(\cdot)$, respectively.

On using the stationarity conditions, corresponding to $u_{k}^{\rm{ch*}}$ and $u_{k}^{\rm{dch*}}$, we obtain (for the sake of simplifying notation, we do not use \say{*} for the dual variables)
\begin{equation}
        \begin{cases}
           {v_{k}^{e}}{(u_{k}^{\rm{ch*}})} + {v_{k}^{r}}{(u_{k}^{\rm{ch*}})} + \big( \lambda_{k}^{\rm{ch}\downarrow}-\lambda_{k}^{\rm{ch}\uparrow} \big) + \frac{\eta^{\rm{ch}}\big( \lambda_{k}^{x\downarrow} -\lambda_{k}^{x\uparrow} \big)}{E} = 0,\\
            {v_{k}^{e}}{(u_{k}^{\rm{dch*}})} + {v_{k}^{r}}{(u_{k}^{\rm{dch*}})} + \big( \lambda_{k}^{\rm{ch}\downarrow}-\lambda_{k}^{\rm{ch}\uparrow} \big) - \frac{\big( \lambda_{k}^{x\downarrow} -\lambda_{k}^{x\uparrow} \big)}{\eta^{\rm{dch}}E} = 0,
        \end{cases}
        \label{eqn:gradient_battery_overall}
    \end{equation}
where $v_{k}^{e}(\cdot)$ and $v_{k}^{r}(\cdot)$ are the subgradients of $f_{k}^{e}$ and $f_{k}^{r}$ with respect to $(\cdot)$, respectively, and we have
\begin{equation}
    \begin{cases}
    {v_{k}^{r}}{(u_{k}^{\rm{ch*}})} &= 2\alpha^{\rm{ch}}u_{k}^{\rm{ch*}} + \beta^{\rm{ch}},\\
    {v_{k}^{r}}{(u_{k}^{\rm{dch*}})} &= 2\alpha^{\rm{dch}}u_{k}^{\rm{dch*}} + \beta^{\rm{dch}},
    \end{cases}
    \label{eqn:gradient_battery_cost}
\end{equation} and
\begin{equation}
    {v_{k}^{e}}{(u_{k}^{\rm{ch*}})} = - {v_{k}^{e}}{(u_{k}^{\rm{dch*}})} = 
    \begin{cases}
    p_{k}, &\quad g_{k} > 0\\
    \delta p_{k}+(1-\delta)s_{k}, &\quad g_{k} = 0\\
    s_{k}, &\quad g_{k} < 0
    \end{cases},
    \label{eqn:gradient_electricity_cost}
\end{equation}
with $0 \leq \delta \leq 1$.\\
\big(Note that the notion of subgradient is used due to the non-differentiability of $f_k^e$ with respect to $u_k^{\rm{ch}}$ and $u_k^{\rm{dch}}$ at $g_k=0$. Further, at the optimal point $v_{k}^{r}(u_{k}^{\rm{ch*}}) \geq0,\,v_{k}^{r}(u_{k}^{\rm{dch*}}) \geq 0,\, v_{k}^{e}(u_{k}^{\rm{ch*}}) >0$\big).

Finally we equate the stationarity conditions in (\ref{eqn:gradient_battery_overall}) to eliminate the coupling w.r.t. the dual variables $\lambda_{k}^{x\downarrow}$ and $\lambda_{k}^{x\uparrow}$, and obtain
\begin{equation}
    \begin{aligned}
        {v_{k}^{e}}{(u_{k}^{\rm{ch*}})} \big( 1-\eta^{\rm{ch}}\eta^{\rm{dch}} \big) + \big({v_{k}^{r}}{(u_{k}^{\rm{ch*}})} + \eta^{\rm{ch}}\eta^{\rm{dch}} {v_{k}^{r}}{(u_{k}^{\rm{dch*}})} \big)\\ + \big( \lambda_{k}^{\rm{ch}\downarrow} - \lambda_{k}^{\rm{ch}\uparrow} \big) + \eta^{\rm{ch}}\eta^{\rm{dch}}\big( \lambda_{k}^{\rm{dch}\downarrow} - \lambda_{k}^{\rm{dch}\uparrow}\big) = 0.
    \end{aligned}
    \label{eqn:stationarity}
\end{equation}
As assumed that the optimal primal and dual solution satisfies the KKT conditions, using complementary slackness we obtain $\lambda_{k}^{\rm{\rm{ch}}\uparrow} = \lambda_{k}^{\rm{\rm{dch}}\uparrow} = 0$, $\lambda_{k}^{\rm{ch}\downarrow} \geq 0$ and $\lambda_{k}^{\rm{dch}\downarrow} \geq 0$. On substituting these values in (\ref{eqn:stationarity}), one can infer that
\begin{equation}
    \begin{aligned}
       {v_{k}^{e}}{(u_{k}^{\rm{ch*}})}\big( 1-\eta^{\rm{ch}}\eta^{\rm{dch}} \big) + \big({v_{k}^{r}}{(u_{k}^{\rm{ch*}})} + \eta^{\rm{ch}}\eta^{\rm{dch}} {v_{k}^{r}}{(u_{k}^{\rm{dch*}})} \big)\\ + \big( \lambda_{k}^{\rm{ch}\downarrow} + \eta^{\rm{ch}}\eta^{\rm{dch}} \lambda_{k}^{\rm{dch}\downarrow} \big) = 0.
    \end{aligned}
\end{equation}
The above equality cannot hold true since the first term is strictly positive, while the second and third term are non-negative. Hence at the optimal point $u_{k}^{\rm{ch}*}u_{k}^{\rm{dch}*} = 0, \forall k \in \mathcal{K}$.
\end{proof}
\vspace{-.5cm}
\section{Results}
We use the residential data of solar houses from AUSGRID dataset \cite{ratnam2017residential}, with the BES as the only controllable load and assume an hourly discretization step, i.e.,  $\Delta t = 1$ and a day-ahead horizon, i.e., $K = 24$. The electricity tariff\footnote{https://www.ausgrid.com.au/Your-energy-use/Meters/Time-of-use-pricing (accessed on April 4, 2021)} corresponds to a working weekday of summer month of Australia. The parameters of the BES are set as $\eta^{\rm{ch}} = \eta^{\rm{dch}} = 0.9$, $\underline{x} = 0.1$, $\overline{x} = 0.9$ and $\alpha^{\rm{ch}} = \alpha^{\rm{dch}} = \beta^{\rm{ch}} = \beta^{\rm{dch}} = 0$.
\begin{figure}[h!]
    \centering
    \includegraphics[width = \linewidth]{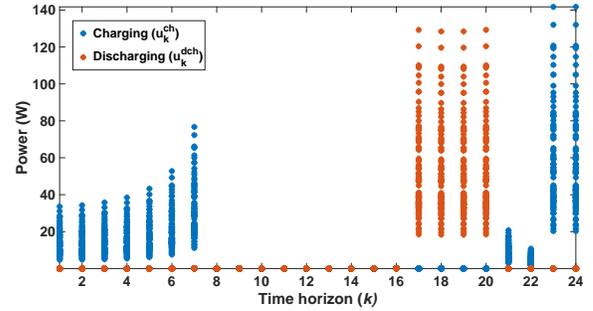}
    \caption{Charging and discharging of BES over 100 houses}
    \label{fig:fig_1}
\end{figure}

The results from Fig. \ref{fig:fig_1}, indicate non-simultaneous charging and discharging of the household battery of 100 houses over a day. 
\vspace{-.5cm}
\bibliographystyle{IEEEtran}
\bibliography{IEEEabrv, IEEETemplate}

\end{document}